\documentclass[a4paper]{mem}
\usepackage{natbib}
\usepackage{graphicx}
\usepackage[a4paper]{hyperref}
\idline{73}{23}
\begin{document}
   \title{Super computers in astrophysics and High Performance simulations of self-gravitating systems
}

   \author{R. Capuzzo-Dolcetta, 
  P. Di Matteo,\\ 
         \and P. Miocchi
         \fnmsep
}

   \offprints{R. Capuzzo-Dolcetta}
\mail{P.le A.Moro 2, I-00185, Roma}

   \institute{Dipartimento di Fisica, Universit\'a di Roma La Sapienza,
P.le A.Moro 2, I-00185, Roma \email{dolcetta@uniroma1.it}\\ 
                     }

   \abstract{
The modern study of the dynamics of stellar systems requires the use of
high-performance computers. Indeed, an accurate modelization of the structure
and evolution of self-gravitating systems like planetary systems, open clusters, globular 
clusters and galaxies imply the evaluation of body-body interaction over the whole 
size of the structure, a task that is computationally very expensive, in particular when
it is performed over long intervals of time. In this report we give a concise overview 
of the main problems of stellar systems simulations and present some exciting  results 
we obtained about the interaction of globular clusters with the parent galaxy.

   }
   \authorrunning{R. Capuzzo--Dolcetta et al.}
   \titlerunning{HPC simulations of self-gravitating systems}
   \maketitle
%

\section{Introduction}


The question "What is a supercomputer?" has a surprisingly trivial answer in the CRAY web site: "A supercomputer is defined simply as the most powerful class of computers at any point in time."
Actually, what modernly distinguishes a supercomputer from normal computers is its more or less complex  {\it architecture} that determines its effective power. Usually, the performances of the high-class processors are quite similar and so producers, in order to increase significantly computing speed, assemble an enormous number of processors linked in the most efficient way. In many cases a relatively cheap solution is that called "Beowulf", i.e. the collection of a certain number of, say, commercial Intel processors linked each other with fast Ethernet. A standard type of "Beowulf" is the so-called Linux cluster.
These systems, are well suited to a wide range of applications  that not require top-performances but, rather,for them suffices a machine whose use is reliable and, possibly, dedicated. Whoever is experienced in numerical simulations knows very well that the largest quantity of time is spent in testing the code in conditions as similar as possible to the ones of the real "long lasting" scientific runs. It is, thus, evident the advantage of having the full compatibility between the {\it testing} machine and the {\it running} machine that is guaranteed by using the same machine. Moreover, in this way all the problems of  large data file remote downloading  is avoided. 
\par\noindent In spite of these considerations, the ``personal'' Beowulf solution is far from being exhaustive. Many scientific problems (I would say the highest-level ones, i.e .the most interesting) require, to be approached with an acceptable level of approximation, higher performances than those reached by "Beowulfs", both in terms of computing speed and memory (Linux  clusters, also when composed by an assembly of bi-processor machines, are limited in the shared memory storage). For this class of problems, where I pose many astrophysical problems, the availability of  huge parallel main frames is essential.  A clear example of this is given by the recent construction of  the Japanese "Earth simulator", a supercomputer explicitly aimed at the detailed simulation of  wheather evolution  over a wide space and time domain with an unprecedented resolution. 
This computer (resulting first for the second consecutive year in the Top 500 international ranking list http://www.top500.org) is composed by 640 nodes of 8 vector processors each giving a total peak power of 35.8 Teraflops/s. This enormous power is largely due to the use of vector processors, which represents a constructive choice different from that, most widely used in U.S., of assembling as many  "commercial" processors as 
possible.
\par In any case, whichever the supercomputers, it is ascertained that science of excellence in general, and astrophysics and cosmology in particular, requires  them. .  

\section {Supercomputing and astrophysics}

Modern astrophysics and cosmology are characterized by dealing with complex problems whose dynamical range covers extended space-time scales.
Some specific topics where this happens are (from small to larger systems): 
\par$\bullet$ stability of solar system,
\par$\bullet$ star formation,
\par$\bullet$ final phases of stellar evolution,
\par$\bullet$ dynamics of galaxies and stellar systems,
\par$\bullet$ physics of AGNs and Quasars (mass accretion onto compact objects),
\par$\bullet$ large scale structure.
\par A correct comprehension of  these topics requires the analysis of the contribution of the many 
processes involved, acting from the micro space-time scales (e.g. atomic and molecular  heating and cooling processes in star formation or relativistic supersonic accretion of matter onto black holes in galactic nulcei activity, etc.) up to the macro structure of the system mainly governed by gravity which is, in addition, 
the main engine of activity also on the micro-scales, due to the irreversible transformation of gravitational energy into thermal. 
\par It is nowadays absolutely clear that firm conclusions about the mentioned open problems cannot be reached by means of simplified models. Consequently, the only way to follow in detail the mutual interaction among phenomena acting on different space-time scales  up to the clear vision of the overall evolution of the system relies on the production of sophisticated numerical modelizations.
Due to the complexity of problems, running these models require indeed high performance computers, i.e. supercomputers. 
\begin{table*}[htb]
\begin{center}
\caption{Stellar systems time scales and other characteristics relevants to the types of simulations needed.The age of the system is $t$. In the last row F-P means Fokker-Planck method, tree-c. means tree-based codes, PM and P3M stand for Particle-Particle and Particle-Particle-Particle-Mesh algorithms, respectively.}
\label{modelli}
\begin{tabular}{|l|c|c|c|} \hline \hline
stellar system:   &open cluster    &glob. clus., gal. nucleus    &galaxy, clus. of gal.\\ \hline
N:      &$<10,000$  &$10^5\div 10^9$  &$>10^{10}$  \\ \hline
gravity:    &newtonian  &newtonian,gen. rel.  &newtonian,gen. rel.  \\ \hline
time scale ordering      &$t_{rel}<t_{cr}<t$  &$t_{cr}<<t_{rel}<t$  &$t_{cr}<<t<t_{rel}$ \\ \hline
regime:     &collisional  &secularly collisional  &collisionless\\ \hline
technique:     &gas+direct N-body  &F-P,direct N-body,tree-c.  &tree-c.,PM,P3M \\ \hline \hline
\end{tabular}
\end{center}
\end{table*}
\subsection{The role of gravity in numerical astrophysics}
By the point of view of computational weight, the most important difference between astrophysical and terrestrial fluid-dynamics is the presence of self-gravity. This implies the substitution  in the equations of motion of the self-gravity body force to the simple fixed  force field. Computationally, this causes an enormous overcharge, due to the long range character of gravity.
This problem is not just a prerogative of collision-dominated systems (fluids, where the relaxation time scale is much shorter than any other dynamical time scale) but typical also of collisionless or partially collisional systems, like galaxies and/or stellar clusters. 
 The long-range behaviour of gravity implies that one cannot neglect the interaction with distant bodies, simply because their number at a fixed distance increases, and this corresponds to an obvious heavy computational weigth at any time step. Things are even more complicated because of the divergence of the newtonian gravitational potential for mutual distances approaching to zero. This is a serious problem whenever fluctuations over the mean field are large in small stellar systems; the short-range behaviour of gravity may 
induce a dramatic reduction of time stepping when a close encounter between two, or more, ``particles'' makes their acceleration to increase up to values such that an acceptable error
in the integration of their trajectories requires a much finer time advancing.
\subsection{From few- to many- body systems}
The study of the stability of solar system is a typical few-body classical gravitational problem.
\par\noindent The number of bodies (planets, asteorids, etc.) is small and  the initial conditions relatively well known so to avoid the problems caused by the double singularity described above. By the way, the secular stability of the solar system is still an unanswered problem due to the intrinsic difficulty in evaluating the role of the resonances that determine the long time behaviour of the system. 
A correct treatment cannot neglect the planet structure, because tides favour resonances among planets and satellites. So, powerful computers are needed to follow the time evolution of a system over very many orbital periods of the components with an accuracy such that one can trust into results that are heavily affected by numerical error accumulation.
 It is intriguing, in this context, that the 3:2 Neptune-Pluto orbital resonance has been found as a result of {\it primitive} supercomputing by \cite{cohen}. 
\par\noindent The actual check of the validity of the results obtained by 
\cite{sussman1}, \cite{laskar} and \cite{sussman2}, that suggest the solar system being chaotic, will probably require the next generation of specifically dedicated supercomputers.
\par Let us say few words about larger N-body problems.
\par\noindent By ``intermediate'' N-body systems we intend stellar systems composed by up to few times $10^6$ stars, i.e. open and globular clusters and clusters of galaxies (in this case, of course, the individual body is a galaxy). The study of the evolution of these systems is a typical ``multiple time scale'' problem, where the {\it fine} grain of the system determines a multiplicity of individual time scales (the two-body fly-by characteristic times $\Delta t_{ij}$) that combine with  the crossing time $t_{cr}$, which, in its turn, depends on the {\it coarse} grain structure. The dynamic range of an accurate time-stepping would span about $5$ orders of magnitude. The use of  time step as the  minimum over this range would cause a CPU time requirement excessively large when the aim is to simulate  the dynamical evolution of the system for a sufficiently long physical time (moreover, the error cumulation would be a great problem). The way commonly adopted to keep accuracy within an acceptable CPU time consumption is the use of individual (and variable in time) time stepping with
accurate sinchronization. 
\par Individual time stepping, inserted in a clever ``tree algorithm'', i.e. such that the treatment of small scale fluctuations of the force field is done by mean of ``local'' direct summation
has a great power in dealing with star cluster dynamics.. Actually, both high accuracy and speed are guaranteed making possible reliable simulations of stellar systems of the size of real globular clusters, with a 1:1 representation of real stars with simulating particles (that means using $N$ up to $10^6$ in the code). 
Such performances are reached, of course, by mean of parallel implementations (see \cite{miocchi}. 
It may be worth stressing that at present,  globular clusters are the {\underline largest} systems that can be studied with a 1:1 representation, i.e. with an almost complete reliability of the results, contrarily to what happens with cosmological and large scale simulations, where the ratio between the number of real objects (stars, galaxies, clusters of galaxies) and simulating particles ($N$) is still very large, making the results, that are also affected by  finite size effects and by the smoothing length of the interaction potential, not 
yet completely reliable and of not easy interpretation.
\section{Globular cluster-galaxy interaction}
\par While the study of the very long-term evolution of a popolous globular cluster is not yet feasible,
because it requires an integration  over a time of the order of the evaporation time, i.e. about one hundred times the 2-body relaxation time, that is $t_{rel}\sim 1,500 t_{cr}$ for a $10^6$ stars cluster, it is 
nowadays possible to examine with good accuracy the evolution of a globular cluster in motion in the galactic field.
\begin{figure*}
   \centering
   \resizebox{\hsize}{!}{\includegraphics[clip=true]{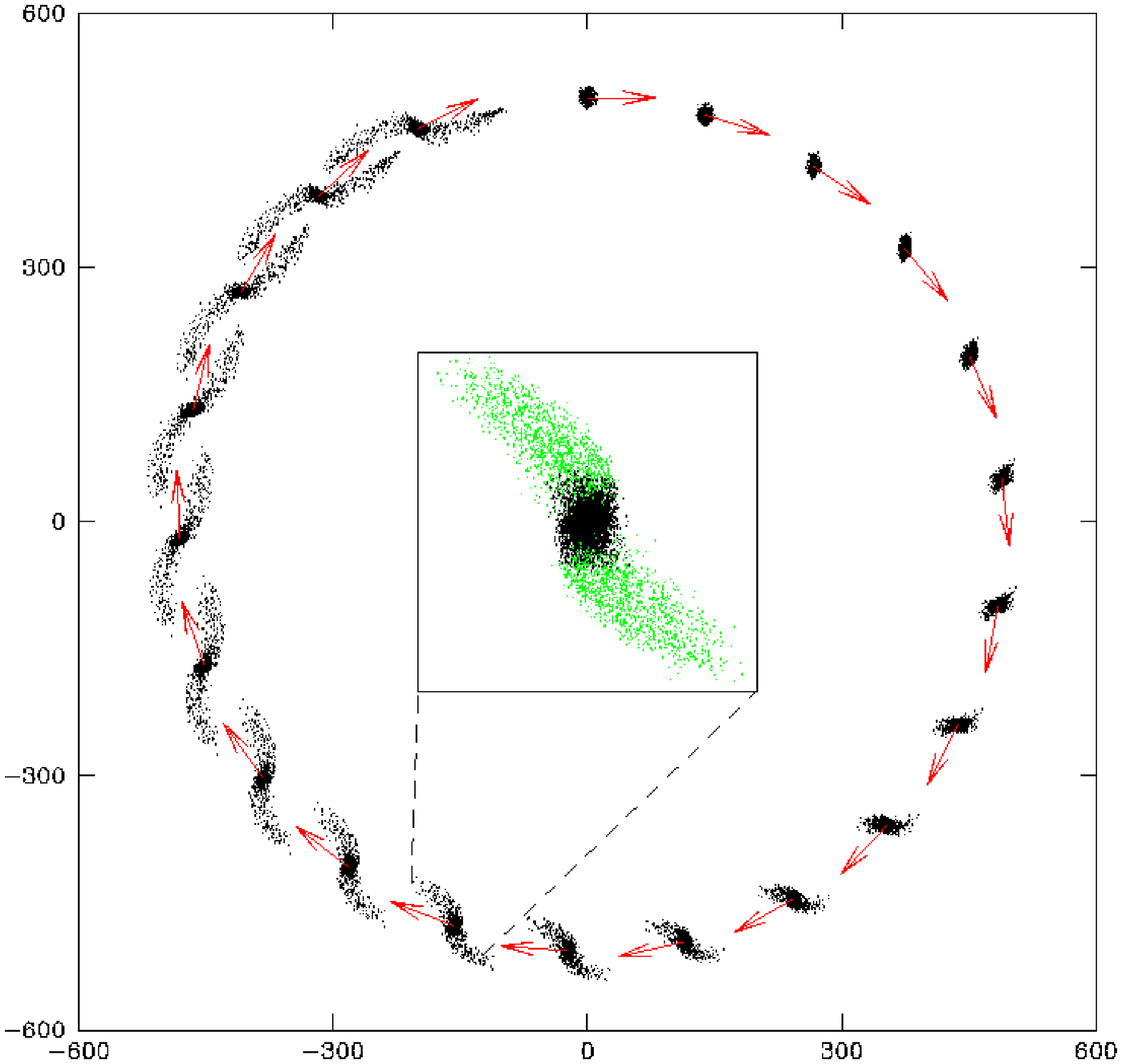}
   \includegraphics[clip=true]{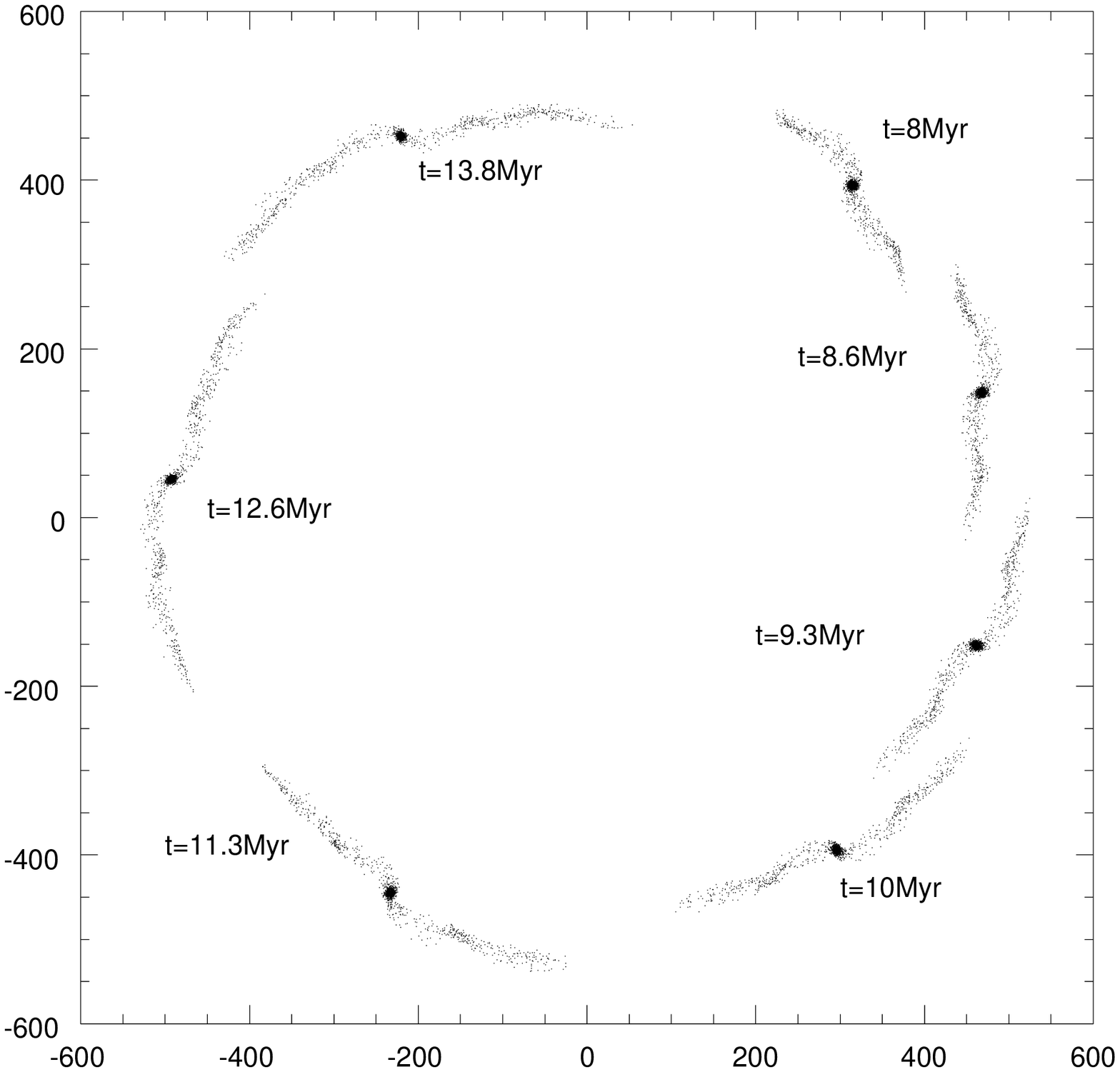}}
     \caption{Left panel: The first period of the clockwise motion of an $3\times 10^5 M_\odot$ globular cluster along a quasi-circular orbit of 500 pc  radius on the y-z plane of the \cite{schwa} triaxial model (the inner box is a zoom of one of the cluster configuration; in green the stars physically belonging to the tails.  Right panel: as in the left panel, but referred to the second  orbital period (snapshots are labeled with time); clumps are evident in the arms}  
        \label{tidaltorque}
    \end{figure*}

 \begin{figure*}
   \centering
   \resizebox{\hsize}{!}{\rotatebox[]{0}{\includegraphics{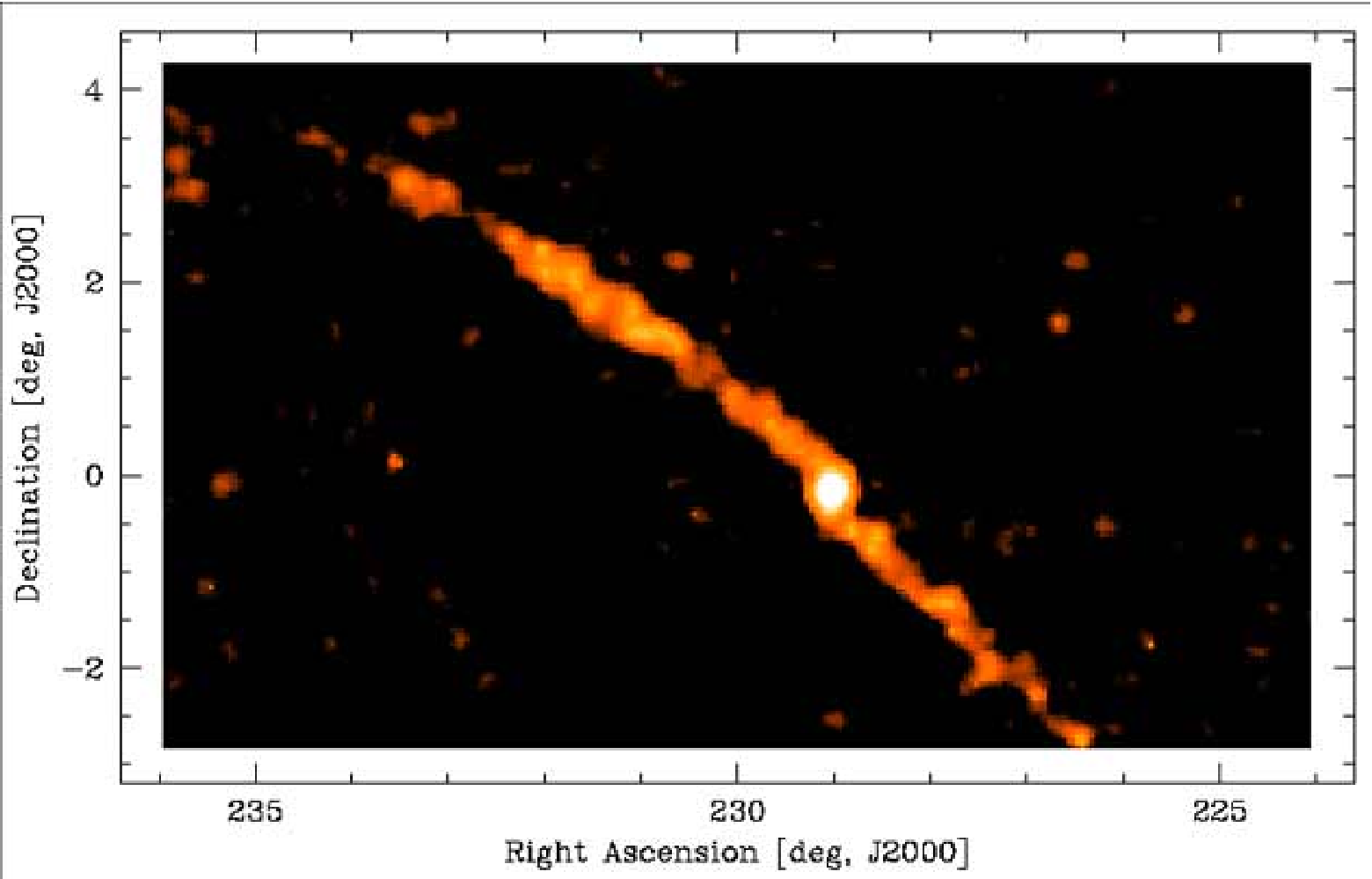}}}
   \caption{ From the SDSS Press release june 3, 2002  (\cite{odenkirchen2}): color-coded map of the distribution of stars emerging from the star cluster Palomar 5 (white blob). The two long tidal tails 
(orange) contain 1.3 times the mass of the cluster and delineate its orbit
around the Milky Way (yellow line).}
              \label{pal5}%
    \end{figure*}
\par\noindent Actually, useful information on the mutual feedback of the cluster with the galactic environment is obtained by simulations which cover an interval of time of 10 orbital times. 
\par\noindent We are presently working in the program of studying the coupled orbital and internal evolution of massive globular clusters in the field of an elliptical galaxy, modeled with a self-consistent triaxial galactic model \cite{dolcetta2}.
The main aims of this program are: 1) to understand why and how tidal tails are formed behind globular cluster moving in the galactic field; 2) to evaluate the amount of mass loss from clusters to the halo-bulge; 3) to check, in a straightforward way, the hypothesis raised time ago by \cite{dolcetta} that a compact object at the galactic centre may accrete a significant quantity of mass by mean of orbitally decayed globular clusters.
\par Here we report just few preliminary results concerning point 1) (they will be extensively presented in \cite{dolcetta1}). Figure \ref{tidaltorque} shows the time evolution of a $3\times 10^5$ solar masses globular cluster made by $N=165,000$ stars whose masses are distributed according to a Salpeter's mass function, with lower and upper mass cutoffs $0.1$ M$_\odot$ and $1.2$ M$_\odot$, respectively. The cluster moves on a quasi-circular orbit of 500 pc radius in the (y,z) plane of the  \cite{schwa} triaxial galactic model of axial ratios 2:1.25:1. It is well evident the development of tidal tails along the orbit, that stabilize their pattern in the direction of the (clockwise) motion after less than one orbital period (about 7 Myr). Note that the front arm points in the direction of motion, exactly as it is seen in the globular cluster Palomar 5 obtained by \cite{odenkirchen2} by mean of wide-field photometric data from the Sloan Digital Sky Survey (SDSS) (Fig. \ref{pal5}).
 Another interesting feature that has an observational counterpart in the same cluster is the formation of clumps along the tails (see Fig. \ref{tidaltorque}, right panel).
Clumps correspond to zones where the tails change directions, probably due to a non-linear cumulative effect of the tidal field. On another side, with regard to the inner stellar distribution, we see that inner relaxation is speeded up by the external field, in the sense that mass segregation is enhanced, thus inducing a decrease of the velocity dispersion in the inner region  in agreement with kinematic data of Palomar 5 
\cite{odenkirchen1} that show that Palomar 5 is a ``cold'' system. 
\par All these results are preliminary and deserve a careful confirm and generalization; by the way we are convinced that these types of numerical simulations are of fundamental importance to give answers to the initial evolution of galaxies as due to mutual feedback among the various primordial stellar populations and also to the still unknown mechanisms of mass accretion onto the AGN engines.

\begin{acknowledgements}

      Part of this work was done thanks to CPU resources in the framework of 
the INAF-CINECA agreement. We thanks  A. Vicari for useful discussions and comments about the orbital evolution of clusters in triaxial galaxies.
\end{acknowledgements}

\bibliographystyle{aa}

\end{document}